\documentclass[twocolumn,prb,prl,nobibnotes,altaffilletter,amsmath,amssymb,amsfonts]{revtex4}
\usepackage[latin1]{inputenc}
\usepackage{graphicx}
\usepackage{amsmath}
\usepackage{latexsym}
\usepackage{epsfig,color}
\usepackage{subfigure}
\usepackage{url}

\begin{document}\title{On polydispersity and the hard-sphere glass transition}

\author{Emanuela Zaccarelli,$^1$ Siobhan M. Liddle$^2$ and Wilson C. K. Poon$^2$}

\affiliation{ $^1$ CNR-ISC Uos Sapienza and  Dipartimento di Fisica, Sapienza Universit\`a di Roma, P.le A. Moro 2, I-00185, Roma, Italy }
\affiliation{ $^2$ SUPA, School of Physics and Astronomy, University of Edinburgh, Mayfield Road, Edinburgh, EH9 3JZ, Scotland}

\date{\today}
\begin{abstract}
We simulate the dynamics of polydisperse hard spheres at high packing fractions, $\phi$, with an experimentally-realistic particle size distribution (PSD) and other commonly-used PSDs such as gaussian or top hat. We find that the mode of kinetic arrest depends on the PSD's shape and not only on its variance. For the experimentally-realistic PSD, the largest particles undergo an ideal glass transition at $\phi\sim 0.588$ while the smallest particles remain mobile. Such species-specific localisation was previously observed only in asymmetric binary mixtures. Our findings suggest that the recent observation of ergodic behavior up to $\phi \sim 0.6$ in a hard-sphere system is not evidence for activated dynamics, but an effect of polydispersity.

\end{abstract}

\maketitle

\textit{Introduction} - 
Despite many decades of experimental, theoretical and simulational effort, the glass transition remains only partially understood. 
The discovery in the 1980s that hard-sphere-like, sterically-stabilised polymethylmethacrylate (PMMA) colloids underwent kinetic arrest \cite{pusey} at a packing fraction of $\phi = \phi_g \approx 0.58$ opened up a fruitful avenue of investigation, because many features of such kinetic arrest in colloids can be mapped onto analogues in atomic and molecular glasses. In particular, hard sphere colloids have become a favourite test bed for mode coupling theory (MCT) \cite{gotzeMCT}. Apart from over-estimating the tendency to vitrify ($\phi_g^{\rm MCT} \approx 0.52$), MCT gives a quantitative account of the main features on approach to arrest, such as a two-step decay in the system's intermediate scattering function (ISF) and power-law dependence of transport coefficients on $|\phi - \phi_g|$ \cite{vanmegen}. 

Such correspondence between MCT and colloidal experiments notwithstanding, doubts have been raised as to whether hard spheres really do undergo a glass transition at $\phi_g \approx 0.58$. Some point to the ease with which monodisperse hard spheres at $\phi \gtrsim 0.58$ crystallise in simulations \cite{torquato} and in microgravity experiments \cite{chaikin} to suggest that the  arrest observed in experiments is due to the inevitable presence in real samples of size polydispersity ($s$, defined as the standard deviation of the size divided by the mean size). However, recent simulations \cite{Zac09a} have shown that particle dynamics near $\phi \approx 0.58$ is nearly invariant for $s \lesssim10\%$. Thus, both polydisperse and monodisperse hard spheres form glasses, only that monodisperse hard spheres are poor glass formers: they crystallise very easily. 

Others suggest that the ideal glass transition in hard spheres is preempted by activated processes not taken into account by MCT. A recent study \cite{brambilla} appears to support this view. Measurements of the ISF of  PMMA colloids over 7 decades in time show that the system remains ergodic at $\phi \gtrsim 0.58$, to at least 0.60, reinforcing the view that activated processes delay the glass transition well beyond MCT; indeed, the suggestion is that there is perhaps no arrest before random close packing $\phi_{\rm rcp} \approx 0.64$. 

The work of Brambilla et al.~\cite{brambilla} has generated significant controversy. Some pointed out their data were largely compatible with MCT if uncertainties in measuring $\phi$ \cite{Poon} were taken into account \cite{fuchscomment}, while others suggested that the large polydispersity ($s > 0.10$) used in \cite{brambilla} to avoid crystallisation could be responsible for the supposed regime of activated dynamics \cite{vanmegencomment}. The authors of~\cite{brambilla} subsequently  simulated  polydisperse hard spheres with a top hat particle size distribution (PSD) and reported relaxation times compatible with experiments, concluding that polydispersity was {\it not} relevant for their findings \cite{cipellettilong,replyvanmegen}. The issue stands unresolved. Its resolution is crucial, since it pertains to the utility or otherwise of a putative model system for glassy arrest.

We have performed extensive simulations of glassy arrest in concentrated, polydisperse hard spheres with PSDs of several shapes, including a PSD measured experimentally by transmission electron microscopy (TEM) for polydisperse PMMA colloids very similar to those used by Brambilla and coworkers \cite{replyvanmegen}.  The distribution is very asymmetric and skewed to the left with an extended tail of small particles, and is well described by a Weibull distribution \cite{langmuir}. We compare the results for this distribution with those that we obtain for gaussian and top hat PSDs with the same variance. We find that the average diffusion coefficient (or equivalently the relaxation time) is independent of the shape of the PSD, but only depends on the variance, and confirm the observations in \cite{brambilla} that no arrest occurs at $\phi \lesssim 0.6$. However, when we focus our attention on subpopulations of particles, we observe a dramatic dependence of dynamics on the PSD. 

For the realistic PSD, relevant for interpreting the data in \cite{brambilla}, we find that the large particles undergo an ideal glass transition, compatible with MCT, at $\phi \sim 0.588$, while the small particles remain mobile. Interestingly, such a localisation transition has been seen before only in asymmetric binary mixtures \cite{dhont,voigtmann,moreno,stars,voigtepl}. Thus, the residual ergodicity reported in \cite{brambilla} at $\phi \gtrsim 0.6$ is {\it not} evidence for activated dynamics but an effect of polydispersity.

\begin{figure}
\includegraphics[width=8.4cm]{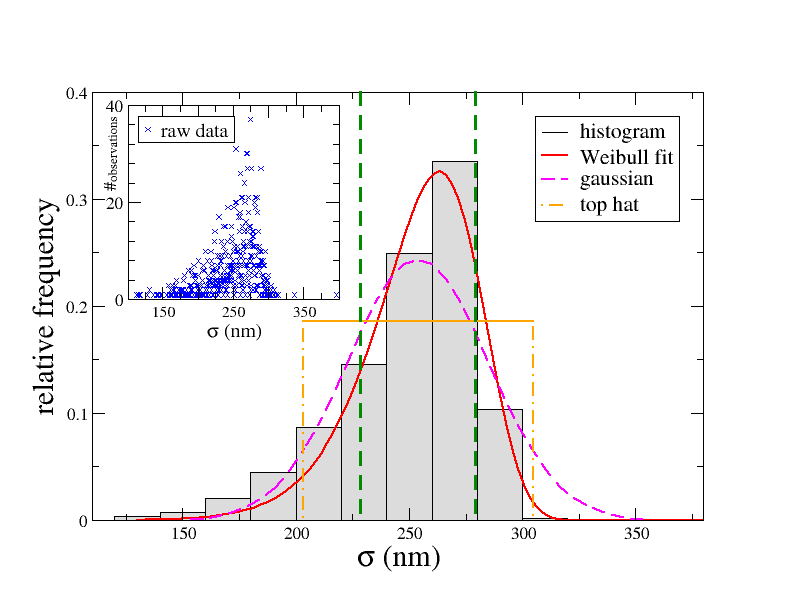}
\caption{Size distribution measured from TEM (histogram) and the fitted Weibull distribution (continuous curve). The average diameter and polydispersity are $\langle\sigma_{\rm TEM}\rangle= 248$nm and $s_{\rm TEM}=12\%$ respectively, while the Weibull fit gives $\langle\sigma_{\rm W}\rangle= 254$nm and $s_{\rm W}=10\%$. Populations of small and large particles are defined as those differing from the average diameter by more than 10\% (left and right of dashed vertical lines).  Also shown are the gaussian (dashed curve) and top hat (dot-dashed curve) PSD with the same polydispersity also used in this work. The latter have been normalised to have equal area to the experimental PSD. Inset: Raw data of PSD as they were used in the simulations.}
\label{fig:exp}
\end{figure}

{\it Methods:}  
We perform event-driven Molecular Dynamics (MD) simulations of hard spheres with different PSDs.  Crucially, this includes a PSD obtained experimentally from PMMA particles synthesised in-house, which were very similar to those used in \cite{brambilla}. The experimental PSD, Fig.~\ref{fig:exp}, was measured from TEM (Phillips CM120 Biotwin)  at $\times$2850, at which the mean particle diameter $\langle\sigma\rangle \approx 40$ pixels. Averaging over $\approx$ 2200 particles gave 
 $\langle\sigma\rangle= 248\pm{4}$nm and $s=12\pm{1}\%$. A Weibull fit well describes the experimental data, Fig.~\ref{fig:exp}. 

We simulate $N=2309$ particles with the experimental PSD, including measurement noise in order to have the most realistic possible representation of the system, Fig.~\ref{fig:exp} inset. We define large and small tail populations as particles with sizes $>(1 + \alpha)\langle \sigma \rangle$ and  $<(1 - \alpha)\langle \sigma \rangle$ respectively. The choice of $\alpha$ does not qualitatively affect our findings (see Supplementary Material). We use $\alpha = 0.1$ to give optimal statistics for the tail populations, which  consist of $\sim 400$ particles each. The unclassified majority ($\sim \frac{2}{3}$) constitute the `average' particles. The large and small populations have average size ratio 1.13:0.8, but extreme size ratios as large as 3 exist in this PSD. For comparison, we also consider $N=2000$ particles taken from gaussian and top hat distributions with the same $\langle \sigma \rangle$ and $s$ as the experimental PSD; the top hat spans sizes in the range $0.8\langle\sigma\rangle$ to $1.2\langle\sigma\rangle$, and its variance is 11.5\% \cite{cipellettilong}. We use units in which the particle mass $m = 1$, average diameter $\bar \sigma = 1$, thermal energy $k_B T = 1$ and time is measured in $\bar \sigma(m/k_B T)^{0.5}$.

In all cases, we first equilibrate the system in the $NVT$ ensemble and then production runs are monitored in the microcanonical ensemble. The mean-squared displacement (MSD) is calculated and from its long-time limit the average self-diffusion coefficient $D$ is extracted.  The averages are performed over different subsets of particles for calculating the small $D_s$ and large $D_l$ diffusivities. At high $\phi$, we monitor aging and distinguish state points showing clear waiting time dependence from equilibrium ones. 
Data collected using the experimental PSD at $\phi\geq 0.59$ have been averaged over ten independent runs. We also calculate the self ISF and the collective ISF at the first peak of the static structure factor and extract corresponding relaxation times, respectively $\tau^{\rm self}$ and $\tau$,  defined as the time where the ISF is at $e^{-1}$.

\textit{Results} - 
\begin{figure}
\includegraphics[width=8.4cm]{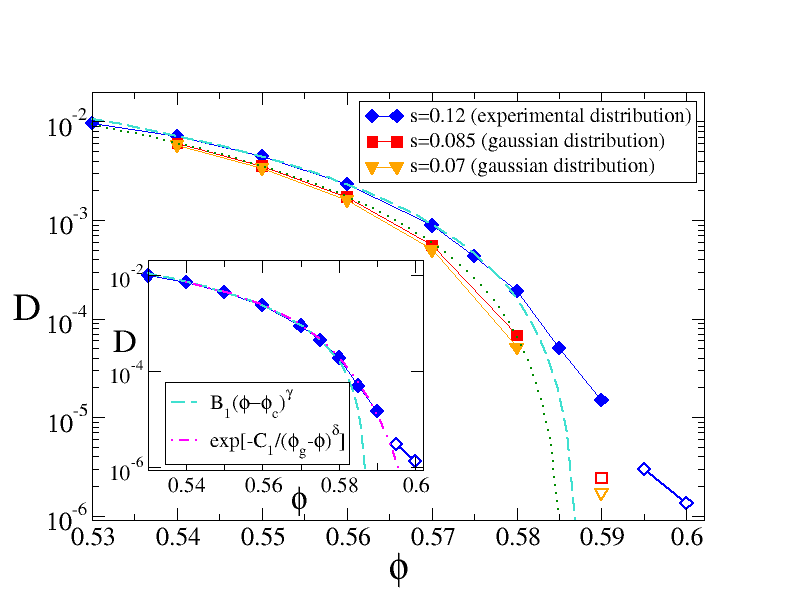}
\caption{Self-diffusion coefficient as a function of $\phi$ for different value of $s$ and different distributions. Closed symbols represent state points that have achieved equilibrium during the simulation run, while open symbols represent state points that show aging. Power-law fit to the data are also shown (dotted line for $s=0.08$, dashed line for $s=0.12$). Inset:
experimental distribution data fitted with an exponential singularity as in \protect\cite{cipellettilong}. We find $\phi_g=0.639$ and $\delta=2.3$.}
\label{fig:diff-s}
\end{figure}
We first monitor $D$ of all particles for the experimental PSD with $s = 0.12$, and compare it to previous results for gaussian PSDs with $s=0.07$ and $0.085$ \cite{Zac09a}, Fig.~\ref{fig:diff-s}. Unlike when $0 < s \lesssim 0.08$ \cite{Zac09a}, the dynamics now become faster with $s$\cite{sear}. More interestingly, the shape of $D(\phi)$ changes upon increasing $s$. For $s \lesssim 0.08$, $D \sim |\phi - \phi_g|^\gamma$ with $\gamma \sim 2.2-2.3$ and $\phi_g \sim 0.585$, consistent with an MCT glass transition \cite{Zac09a}, with aging in states above $\phi_g \sim 0.585$. On the other hand for $s=0.12$, the decrease of $D(\phi)$ becomes much less pronounced at high $\phi$, a power-law behavior fails to account of all fluid state points. Indeed, the system can still be equilibrated  for $\phi \sim 0.59$, where a strictly finite value of $D$ is found, well above that predicted by a  power law fit.  A better description of the data is given by an exponential singularity $D\sim \exp\left[1/(\phi-\phi_g)^{\delta}\right]$ (cf.~Fig.~\ref{fig:diff-s} inset), where $\phi_g \approx \phi_{\rm rcp}$ and  $\delta\approx 2.3$, suggesting an approximate double exponential singularity. These results are in full agreement with previous experiments for a similar system \cite{brambilla}.

\begin{figure}
\includegraphics[width=8.4cm]{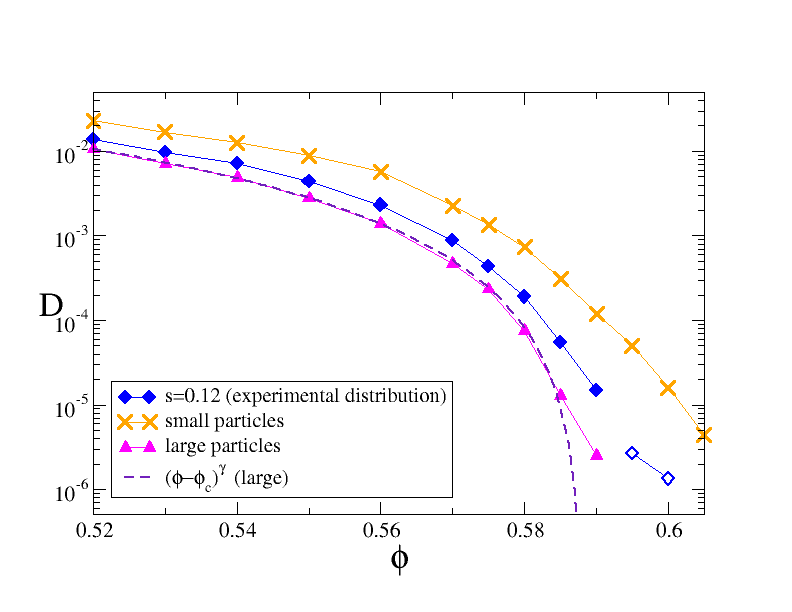}
\caption{Self-diffusivity as a function of $\phi$ for the experimental distribution, differentiating small and large particles from the average. A power-law fit (dashed line) with $\gamma\sim2.3$ and $\phi_c^{l}\sim 0.588$ describes well the large particles data, while small particles remain mobile at all investigated $\phi$.}
\label{fig:diff-small-large}
\end{figure}

We next consider $D$ for the tails of the experimental PSD, Fig.~\ref{fig:diff-small-large}. At $\phi =0.59$, there is already a clear dynamical separation between the small and large particles, with $D_s \sim 10^2 \times D_{\l}$ , suggesting a scenario previously observed only in binary asymmetric mixtures \cite{dhont,voigtmann,moreno,stars,voigtepl}, 
where small particles remain mobile in a matrix of arrested large ones. Most importantly, $D_\ell \sim |\phi - \phi_g^{\l} |^\gamma$ with $\phi_g^{\l} \sim 0.588$ and $\gamma\sim 2.3$, i.e.~the large particles in the remainder of the system retain ideal hard-sphere glassy behaviour. However, the system as a whole shows significant deviations from this behavior. In particular, we do not observe arrest of the small sub-population up to the largest $\phi=0.605$. These motile smaller particles provide residual ergodicity well above $\phi = 0.58$. This residual ergodicity is {\it not} the result of activated processes; rather, it is a direct consequence of polydispersity. 

\begin{figure}
\includegraphics[width=8.4cm]{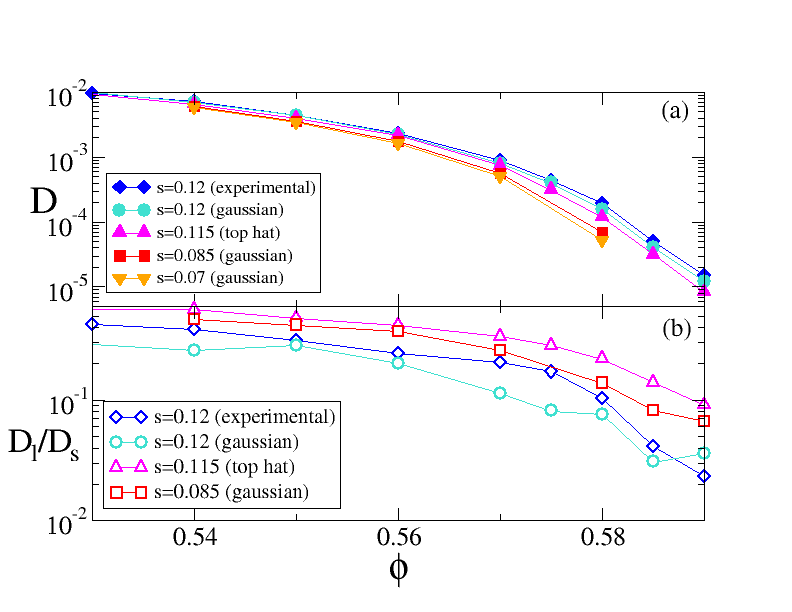}
\caption{(a) Average self diffusion coefficient and (b) ratio of the large to small diffusivity as functions of $\phi$: experimental, $s=0.12$ (diamonds);  gaussian, $s=0.12$ (circles); top hat, $s=0.115$ (up triangles); gaussian, $s=0.085$ (squares) and $s=0.07$ (down triangles).}
\label{fig:diff-shapes}
\end{figure}

It is important to ask whether these findings depend specifically on the PSD, and if so whether they are mostly determined by $s$ or by the PSD's whole shape. We therefore repeat the simulations for state points where full equilibrium could be achieved using a gaussian distribution with $s=0.12$ and a top hat distribution with $s=0.115$, Fig.~\ref{fig:exp}. Figure~\ref{fig:diff-shapes}(a) shows that averaged over the whole system, $D(\phi)$ is essentially independent of the shape of the PSD, but depends only on its variance. 

However, important differences emerge between the top hat and the other two PSDs at comparable $s$ when we analyse the contributions from small and large particles, plotted in Fig.~\ref{fig:diff-shapes}(b) as the diffusivity ratio $D_l/D_s$. As seen before, $D_l$ has decreased to $\sim 10^{-2} D_s$ at $\phi = 0.59$ for the experimental PSD. With the top-hat PSD, this diffusivity ratio is $\approx 10$ times less extreme, i.e., selective localisation of the larger particles is a much weaker effect. Presumably, this is because defining sub-populations of small and large particles makes little sense in a uniform distribution compared to the same exercise in strongly-peaked PSDs. In addition, unlike the experimental PSD, the sub-population of large particles in the top-hat PSD does not clearly show an ideal glass transition (see Supplementary Material). Thus, a uniform distribution with the same $s$ as the experimental PSD does {\it not} reproduce key qualitative features of the microscopic dynamics of a system with the experimental PSD.  

Note from  Fig.~\ref{fig:diff-shapes}(b) that using a (peaked) gaussian PSD largely reproduces the behavior of the experimental PSD, especially at $\phi \gtrsim 0.58$. Interestingly, there is residual diffusivity in all three PSDs at $\phi \gtrsim 0.59$. Even in the top hat, this is due to a gradation of dynamics from large to small particles  (see Supplementary Material), despite the absence of partial arrest. Figure~\ref{fig:diff-shapes}(b) also shows that lowering $s$ decreases the difference between $D_s$ and $D_{\l}$. Thus, the low-$s$ colloids used by Pusey and van Megen \cite{pusey,vanmegen,martinez} should show ideal arrest of all particles simultaneously at  $\phi\sim 0.58-0.59$ with no partial localisation.

\begin{figure}
\includegraphics[width=8.4cm]{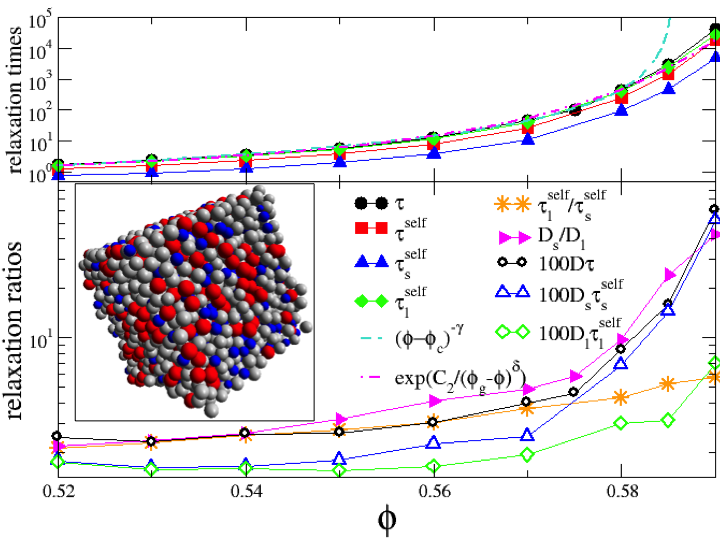}
\caption{Relaxation times as a function of $\phi$ for the experimental size distribution.
Top panel: collective $\tau$, self $\tau^{\rm self}$, self for small $\tau^{\rm self}_s$ and large $\tau^{\rm self}_l$ particles; Bottom panel: ratios of large and small particles transport coefficients as a function of $\phi$. Note that $D\tau$, $D_s\tau^{\rm self}_s$ and $D_l\tau^{\rm self}_l$ have been multiplied by a factor 100 to be on the same scale. Inset: a snapshot of the system at $\phi= 0.60$.  Particles, represented at their full sizes, are colour coded: blue (small), red (large) and grey (intermediate-sized). The small particles can be seen as moving in a random, porous matrix of larger particles.}
\label{fig:tau}
\end{figure}

Finally, we consider collective and self relaxation times, averaged over all particles, $(\tau,\tau^{\rm self})$, and averaged over the small and large sub-populations, $(\tau_s, \tau_s^{\rm self})$ and $(\tau_{\l}, \tau_{\l}^{\rm self})$. Data for the experimental PSD are shown in Fig.~\ref{fig:tau} (top). No significant difference is observed for the form of the dependence on $\phi$ for the other two PSDs (data not shown). Significantly, there appears to be no decoupling between self and collective relaxation: $\tau(\phi)$ and $\tau^{\rm self}(\phi)$ remain parallel at all $\phi$ studied here. Such decoupling might have been expected from the diffusivities of these sub-populations, Fig.~\ref{fig:diff-shapes}(b). However, studies of binary mixtures \cite{morenoJCP,voigtmann} have shown that a full decoupling between self and collective dynamics only occurs for size ratios $\gtrsim 5$, beyond the extremes of size disparity that are present in our experimental PSD, Fig.~\ref{fig:exp}. Quantitatively, an exponential function of the form $\exp\left[1/(\phi-\phi_g)^{\delta}\right]$ fits the $\phi$-dependence of these relaxation times rather than a power law, with $\phi_g \approx \phi_{\rm rcp}$  and $\delta\approx 2.34$. 

Turning to the ratios of partial transport coefficients, the variation of $D_s/D_l$ with $\phi$ exceeds that of $\tau^{\rm self}_l/\tau^{\rm self}_s$ by roughly one order of magnitude,  Fig.~\ref{fig:tau} (bottom). As $D_{\l}$ begins to drop precipitously relative to $D_s$ at high $\phi$, $\tau_{\l}^{\rm self}$ fails to rise in proportional relative to its `small' counterpart, $\tau_s^{\rm self}$. If we take $\tau^{\rm self}$ as a surrogate for viscosity, then the lack of scaling between $D$ and  $\tau^{\rm self}$  can be seen as a manifestation of the violation of the Stokes-Einstein relation (SER). 

SER violation is ubiquitous in atomic and molecular glass formers \cite{SEviolation,SEviolation2}, where the product $D\tau$ increases on approaching the glass transition. This is commonly interpreted as evidence for increasing spatial heterogeneities in the dynamics. In our case, there is a far sharper rise in $D_s\tau^{\rm self}_s$ with $\phi$ than in $D_l\tau^{\rm self}_l$. This indicates a high degree of dynamic heterogeneity for the small sub-population. A system snapshot, Fig.~\ref{fig:tau}, and a close inspection of partial structure factors (not shown), reveal that large and small particles are randomly distributed. Thus, our polydisperse system at $\phi\gtrsim 0.59$ can be conceptualised as a fluid of small spheres in a porous matrix of larger spheres close to the glass transition \cite{krakoviack,kurdizim,kim}, where a high degree of dynamic heterogeneity in the mobile (= small in our case) particles has been observed \cite{kurdizim2}.

\textit{Conclusions} We have shown that polydispersity has highly non-trivial effects on the hard sphere glass transition. In particular, experimentally realistic, peaked PSDs can preempt an ideal glass transition at $\phi\sim 0.58$ by inducing partial arrest of the largest particles, while the smaller ones remaining  largely diffusive. Such partial arrest was {\it not} observed in the top-hat PSD. In all cases, differential mobility of the particles induces residual ergodicity above $\phi = 0.58$, but this effect that should not be confused with generic activated processes.

 Our findings are compatible with the view that moderately polydisperse hard spheres, $s\sim 5-6\%$, can function as a reference system for glassy arrest.
On the other hand,  hard spheres with larger polydispersity, $s \gtrsim 0.10$, at large packing fractions, $\phi\gtrsim0.59$, behave as {\it hierarchical fluids} as far as dynamics is concerned, where several particle populations, with large dynamic heterogeneities, are present. It remains a challenge to understand the microscopic mechanisms operative in such systems; this will be addressed in future work.

\textit{Acknowledgements} We thank A. Schofield for particles, and S. Egelhaaf, M. Laurati, V. Martinez, E. Sanz and C. Valeriani for discussions.
We acknowledge support from ITN-234810-COMPLOIDS. EZ acknowledges support from MIUR-FIRB ANISOFT (RBFR125H0M). SML was funded by an EPSRC studentship. WCKP and part of EZ's visit to Edinburgh were funded by EPSRC grant EP/J007404/1. TEM was performed in the Wellcome Trust Centre (UoE). Simulations were performed using resources provided by the Edinburgh Compute and Data Facility (ECDF) (\url{http://www.ecdf.ed.ac.uk/}).

\renewcommand{\thefigure}{S\arabic{figure}}
 \renewcommand{\thetable}{S\arabic{table}}
\setcounter{figure}{0}

\section*{Supplementary Material}
\begin{figure}[h]
       \includegraphics[width=8.4cm]{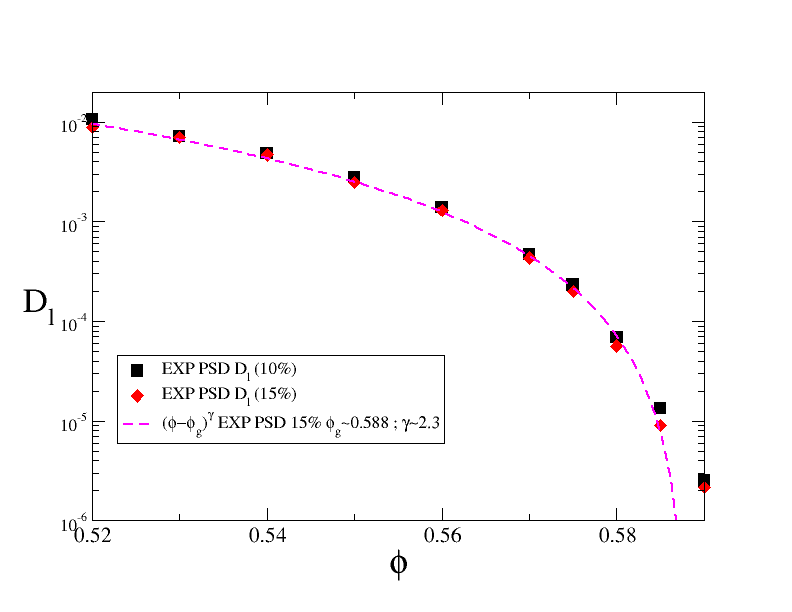}
\caption{Diffusion coefficients of the large particles $D_l$, defined using two different values of $\alpha$, 
for the experimental PSD. Both data sets can be described by a very similar power-law fit yielding the same estimate for $\phi_g^{\l}$ and exponent $\gamma$.}
 \label{supp1}
\end{figure}

We first show that the choice of $\alpha$ does not significantly affect our results. In the main text, we defined the large and small sub-populations as those particles with sizes $>(1 + \alpha)\langle \sigma \rangle$ and  $<(1 - \alpha)\langle \sigma \rangle$ respectively. The diffusion coefficient of the large particles, $D_l$, defined using $\alpha = 0.1$ is shown in Fig.~\ref{supp1} as a function of $\phi$. These are the data already shown in Fig.~3 in the main text. In the same figure we plot $D_(\phi)$ calculated using $\alpha = 0.15$. While these data suffer from poorer statistics compared to the case of $\alpha = 0.1$, they can still be well described by the same power law fitted to the $\alpha = 0.1$ data in the main text, viz., $D_\ell \sim |\phi - \phi_g^{\l} |^\gamma$ with $\phi_g^{\l} \sim 0.588$ and $\gamma\sim 2.3$. Thus, it makes no material difference whether we choose $\alpha = 0.1$ or $\alpha = 0.15$.

\begin{figure}[h]
       \includegraphics[width=8.4cm]{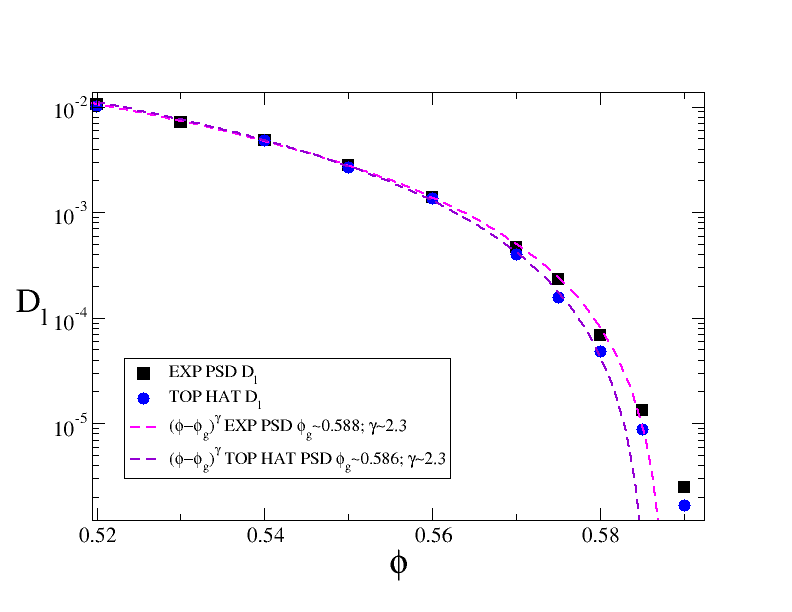}
\caption{Diffusion coefficients of the large particles $D_l$ for the top hat distribution and 
for the experimental PSD. Applying a power-law fit to the former set of data yields an anticipated transition at $\phi_g^{\l} \sim 0.585$, which clearly misses the last equilibrium data point in the simulations.}
 \label{supp2}
\end{figure}

Next we show that a power law fit, associated to
an MCT-type ideal glass transition, does not equally well describe the behavior of the large sub-population in the top-hat PSD. Fig.~\ref{supp2}~compares $D_l(\phi)$ for the experimental and top-hat PSDs defined using $\alpha = 0.1$. As we have already shown in the main text, $D_\ell \sim |\phi - \phi_g^{\l} |^\gamma$ with $\phi_g^{\l} \sim 0.588$ and $\gamma\sim 2.3$ describes the experimental PSD data well for all but the last data point, which showed ageing and therefore must be above any putative $\phi_g$. Fitting the same functional form to the top-hat data returns the same $\gamma$, but a transition point of $\phi_g^{\l} \sim 0.585$. The measured $D_l$ for the top-hat PSD at this $\phi$ is substantial, and the system at this $\phi$ does not show ageing. 

\begin{figure}[h]
       \includegraphics[width=8.4cm]{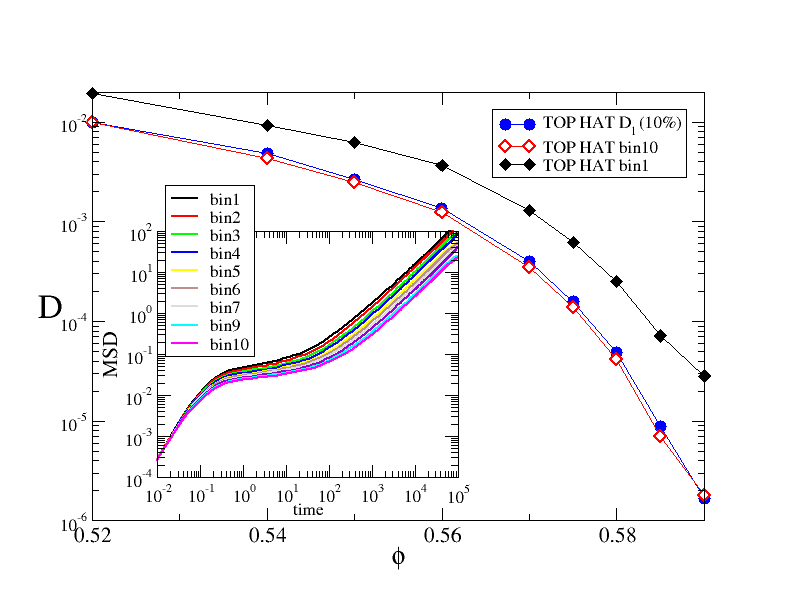}
\caption{Diffusion coefficients of the large particles $D_l$ for the top hat distribution, compared to those calculated in a 10-bin analysis of the distributions. Bin1 contains the smallest particles, while bin10 contains the largest ones.
Inset: MSD versus time for the particles in different bins at $\phi=058$.}
 \label{supp3}
\end{figure}

Thus, we conclude that we cannot identify a proper arrest transition of the sub-population of large particles for the top-hat PSD. This conclusion is insensitive to the precise definition of `large'. Figure~\ref{supp3} compares the $D_l(\phi)$ data  for the top-hat PSD already given in Fig.~\ref{supp2} with those calculated for the particles in the largest bin in a classification of the top-hat PSD into 10 equal bins. There is no material difference in the results. 

Finally, we show that the residual motions detected in the top-hat PSD at high $\phi$ are still due to a gradation of dynamics, from slower, larger particles to faster, smaller particles, although this is not as extreme as for the experimentally realistic PSD.  Fig.~\ref{supp3} shows this in two ways. In the main figure, we compare the diffusivities of the particles in the smallest and largest bin in a 10-bin division of the top-hat distribution as a function of $\phi$. The inset shows the mean squared displacements (MSD) as a function of time for all 10 bins at $\phi=0.58$. Note that all these bins have equal populations, weighting equally on the average of the total $D$, resulting in a smearing of differences that are, by contrast, enhanced by the presence of peaks and tails in more realistic PSDs.

\end{document}